\documentclass[]{acm_sen_article}

\usepackage{lipsum}
\usepackage{flushend}

\usepackage{graphicx} 
\PassOptionsToPackage{hyphens}{url}
\usepackage{hyperref}
\usepackage{xspace}
\usepackage[numbers]{natbib}
\usepackage{balance}
\usepackage{xcolor}
\hypersetup{
  colorlinks=false,
  pdfborder={0 0 0}
}

\usepackage{fontawesome5}
\usepackage{framed}
\definecolor{formalshadelight}{RGB}{242,242,242}
\definecolor{formalshadedark}{RGB}{166,166,166}

\usepackage{csquotes}

\begin{document}

\title{The State of Peer Review in Empirical Software Engineering: A Community Survey on Review Load, Quality, and GenAI Use}

\subtitle{ACM SIGSOFT SEN-ESE Column}

\numberofauthors{2} % number of author blocks
\author{
Justus Bogner\\
       \affaddr{Vrije Universiteit Amsterdam}\\
       \affaddr{The Netherlands}\\
       \email{j.bogner@vu.nl}\and
       Roberto Verdecchia\\
       \affaddr{University of Florence}\\
       \affaddr{Italy}\\
       \email{roberto.verdecchia@unifi.it}
}

\maketitle

% For page numbers at the bottom
\pagenumbering{arabic}

\begin{abstract}
The scientific peer review system has been slowly deteriorating over the last years, and not just within empirical software engineering (ESE) research.
Increased submission numbers, high workload, and the rise of generative AI use with all its associated issues have made many cracks in the system more visible.
To get a better understanding of the current state of peer review in the ESE community, we conducted a questionnaire survey, which accumulated 120 responses.
We report on (i) the perceived review load of community members, (ii) review quality perception as well as frequent challenges for and issues with reviews, (iii) the use of LLM-based tools in the reviewing process, and (iv) the community’s suggestions for improving the
peer review system.
We hope that these community opinions can facilitate more evidence-based discussions about how people want to see the review system change for the better.
\end{abstract}

\section{Introduction}
\label{sec:intro}

For decades, peer review has served as a cornerstone of quality assurance in scientific research, providing a mechanism for submitted manuscripts to be evaluated by domain experts before publication~\cite{Aczel2025}.
The empirical software engineering~(ESE) community is no different: venues such as the International Conference on Software Engineering~(ICSE), the Empirical Software Engineering journal (EMSE), the International Symposium on Empirical Software Engineering and Measurement~(ESEM), and the International Conference on Evaluation and Assessment in Software Engineering~(EASE) all rely on unpaid volunteers as peer reviewers to maintain scientific standards and to guide researchers in improving their work.

Despite its central role, the peer review system has come under heavy load and increasing criticism in recent years.
The global growth in manuscript submissions has far outpaced the supply of qualified reviewers~\cite{Shah2022, Aczel2025}, leading to reviewer fatigue, superficial evaluations, and growing rates of review decline~\cite{Alchokr2022, Soldani2020}.
Several studies have revealed that reviews are often unreliable and inconsistent, e.g., the famous NeurIPS'14 experiment~\cite{Cortes2021} and its replication in 2021~\cite{Beygelzimer2023}, which showed that a substantial fraction of accepted papers would have been rejected had they been reviewed by a different panel (\enquote{the two committees disagree on their accept/reject recommendations for 23\% of the papers and that, consistent with the results from 2014, approximately half of the list of accepted papers would change if the review process were randomly rerun}~\cite{Beygelzimer2023}).
The review process can also be prone to systematic biases related to author prestige, gender, and institutional affiliation~\cite{Aczel2025, Prechelt2018}, and junior researchers may face disproportionate barriers when participating in the reviewing ecosystem~\cite{Alchokr2022}.
Within the ESE community in particular, surveys of authors and program committee (PC) members have revealed that roughly one third of all reviews are perceived as useless or misleading~\cite{Prechelt2018, Ernst2021}.

In response to these challenges, the community has proposed a wide range of remedies:
adopting double-anonymous or open reviewing models to reduce bias~\cite{Prechelt2018},
introducing empirical standards to make review criteria explicit and
consistent~\cite{Arshad2021}, encouraging structured review forms that guide
reviewers toward the most important dimensions of a
submission~\cite{Arshad2021, Aczel2025}, and investing in reviewer
training programs like junior PCs~\cite{Ernst2021}.
While several of these initiatives show promise, the scientific community has not yet reached broad consensus on their effectiveness~\cite{Shah2022, Soldani2020}.
 
The rapid adoption of generative artificial intelligence~(GenAI), in particular of large language models~(LLMs), has considerably worsened the state of peer review~\cite{Aczel2025}.
On one hand, LLM-increased writing productivity has led to substantially more submissions for the peer review system to deal with~\cite{Kusumegi2025}.
On the other hand, reviewers have also started to make use of LLMs, often with detrimental effects.
For example, an analysis of major machine-learning conferences estimates that between 6.5\% and 16.9\% of submitted review text was substantially generated or modified by LLMs~\cite{Liang2024}, raising urgent concerns about the integrity, accountability, and depth of expert evaluation.
These AI-assisted reviews tended to be more shallow, were more prevalent among low-confidence reviewers submitting close to the deadline, and may introduce novel forms of bias that are difficult to detect~\cite{Liang2024}.
While the broader scientific community is still debating how to regulate or productively integrate LLMs into scholarly workflows~\cite{Aczel2025}, the ESE community has yet to systematically characterize the extent and impact of LLM use in its own review processes.
 
To better understand the current state of peer review in ESE research, we conducted a questionnaire survey~\cite{Wagner2020} targeting reviewers across the major ESE venues.
Concretely, we investigate (i)~the perceived review load of community members, (ii)~review quality perception as well as frequent challenges for and issues with reviews, (iii)~the use
of LLM-based tools in the reviewing process, and (iv)~the community's suggestions for improving the peer review system.

\section{Study Design}
The questionnaire started with some basic demographic questions like participant role or location.
It then covered questions about review load, review quality and process, LLMs and reviewing, and ended with suggestions to improve the current state of reviewing.
Of the 22 questions, only the final 2 were open free-text questions, with the rest being single-choice or multiple-choice questions.
Most of these also had an \enquote{other} option to allow adding custom answers.
The initial questionnaire was piloted with four experienced ESE researchers, who provided feedback and suggestions for improvement.
As a result, several questions and answer options were refined, and a few questions were dropped to keep the answer time closer to 5 min.

Our requirements for participation were that people needed to have acted as an official peer reviewer for a scientific paper at least once in the past 12 months and that a decent part of their research and reviewing activity was in the ESE area.
We advertised the survey via our personal contacts and social media, but also by emailing the PC of the 2026 editions of the ESEM, ICSE, and FSE conferences.
Responses were collected fully anonymously, without any means to identify participants.
Before starting, participants had to consent to the described data collection and usage policy.

After closing the survey, we downloaded, cleaned, and transformed responses for the analysis.
Questions with multiple answers per participant were moved into their own spreadsheet tab to make aggregation easier.
For free-text questions, we used thematic analysis to label and organize the answers.
For transparency and replicability, we publish the survey data online.\footnote{\url{https://doi.org/10.5281/zenodo.20495019}}

\section{Participant Demographics}
\begin{figure}
    \centering
    \includegraphics[width=\linewidth]{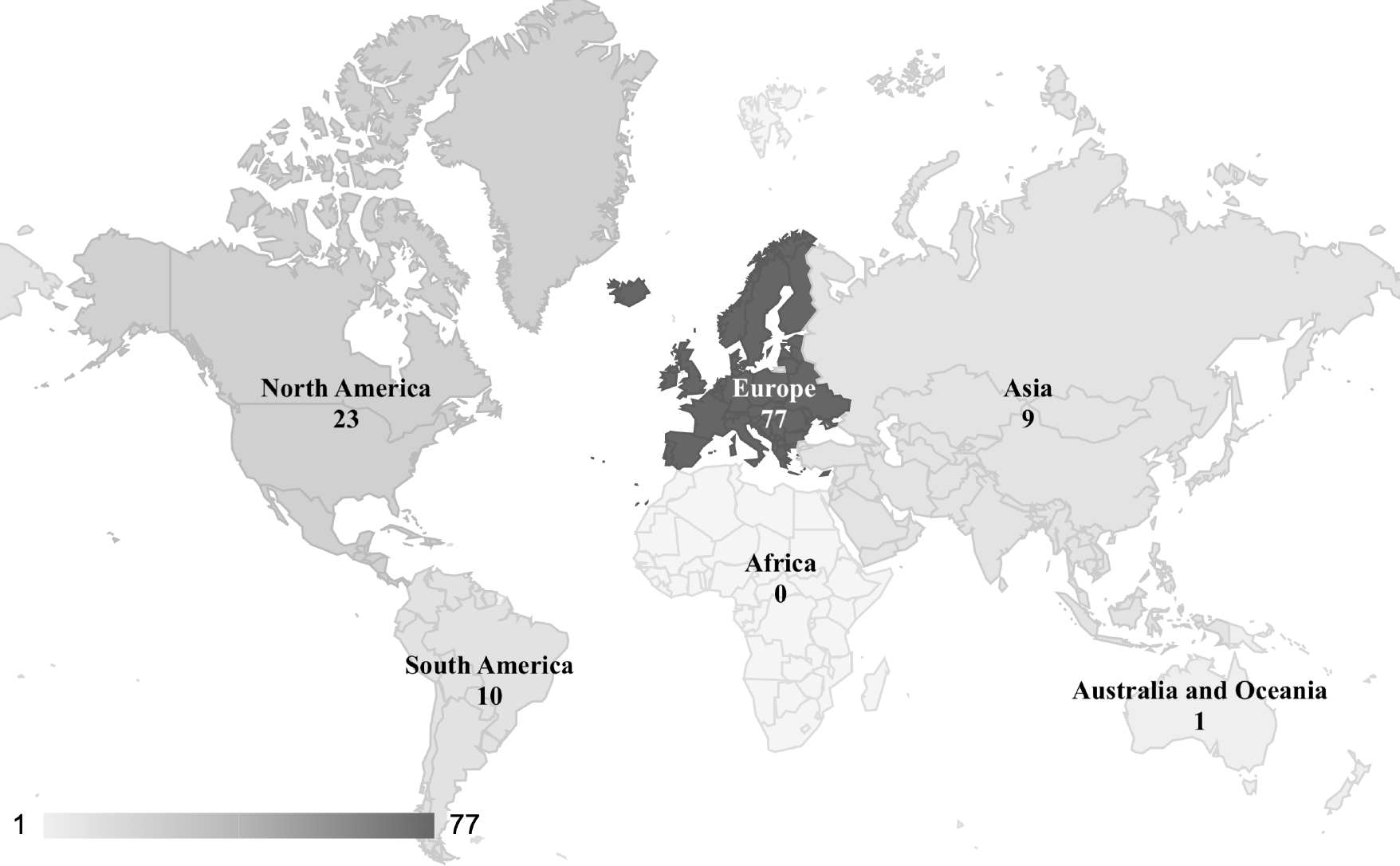}
    \caption{Geographical distribution of the 120 participants}
    \label{fig:continents}
\end{figure}

The vast majority of the 120 survey participants held a tenured academic position (81), followed by postdocs (13), other non-tenured academic roles (12), industry practitioners or researchers (10), and PhD students (4).
These roles align with the experienced nature of our participants: most have worked in ESE research for a consolidated period of 11-20~years (59).
Several respondents have dedicated an even larger part of their life to the topic, from 21-30~years (20) to more than 30~years (3).
Junior ESE researchers were less present in our sample and ranged from fewer than 3~years (1) to 3-5~years (12) and 6-10~years (25).
Regarding geolocations, an overview is provided in Fig.~\ref{fig:continents}.
Participants were predominantly located in Europe (77), with North America taking second place (23).
A minor portion of respondents worked in South America (10), Asia (9), or Australia and Oceania (1).
No ESE researchers from the African continent participated.
In summary, \textbf{most survey respondents were seasoned ESE researchers from European or North American institutions}, which needs to be considered for interpreting the results.

\section{Reviewer Load}
\label{sec:load}
\begin{figure}[h]
    \centering
    \includegraphics[width=\linewidth]{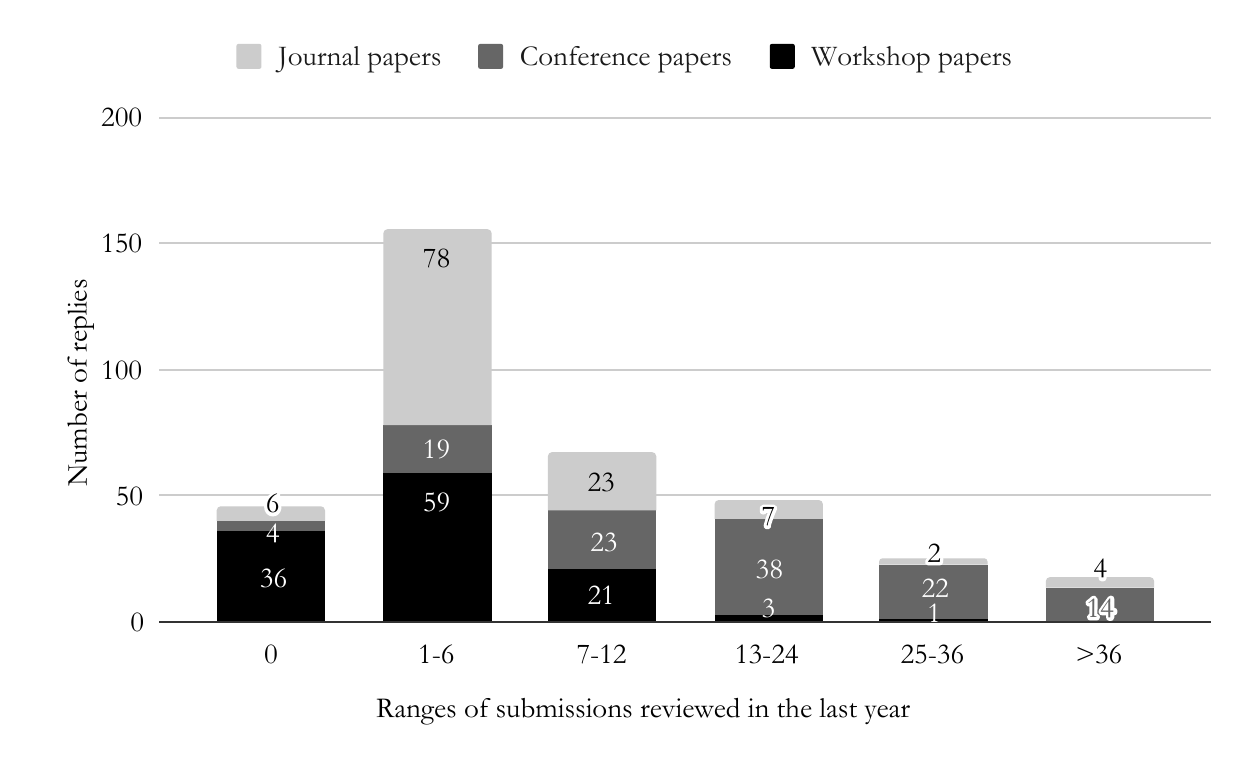}
    \caption{Distribution of ESE reviewing effort across workshops, conferences, and journals in the last year}
    \label{fig:load}
\end{figure}

To gain insights into the ESE reviewing load, our 120 participants had to rate their perceived review load in the last year on a 5-point ordinal scale (\enquote{very low} to \enquote{very high}).
A two-thirds majority perceived their review load as high (51) or very high (29), with a median rating of 4 and a mean of 3.88.
Conversely, 36 people selected \enquote{medium}, with only 4 people reporting a low review load and no one choosing \enquote{very low}.
To compare this perception with performed reviews, we asked participants how many papers they roughly reviewed in the last year across workshops, conferences, and journals.
Reviews of revisions or meta-reviews were not included.
An overview of the responses is documented in Fig.~\ref{fig:load}. 

Journal and workshop reviewing roughly followed the same distribution, with most participants reviewing about 1-6 papers for each of the two venue types.
One difference between the two was that 30\% of respondents did not participate in any workshop PC (36), but only 5\% did not review a single journal paper (6).
However, the picture changes completely for conferences, with a pronounced shift to the right.
The most reported category was 13-24 conference reviews (38), with 22 participants reporting 25-36 reviews and 14 even reporting more than 36 reviews, i.e., more than 3 conference reviews per month.
By combining all venue types, we estimate the average number of reviews per person and year to be somewhere between 25 and 32, of which roughly two-thirds are spent on conferences.
That means 2-3 reviews per person per month, but with a fairly uneven distribution.

When grouping by demographics, we observe that seniority aligns well with increased review load, suggesting that at least this aspect of the system seems to work.
The small sample of PhD students and industry participants experienced a lower reviewing load, with mostly conference reviews.
The remaining categories showcase higher review loads with broadly similar distributions: conferences dominate, journals are moderate, workshops are light.
Tenured positions also display a considerably heavier conference review load, with 12 of the 14 responses for more than 36 reviews last year being tenured positions.
This seems to indicate that, on average, senior community members are indeed pulling their weight.
A rough guideline to ensure fairness in review load that is mostly accepted within the ESE community suggests reviewing three papers for each submitted one~\cite{Ernst2021,Parra2026}, with junior members like PhD students being exempted.
We asked participants how this ratio worked out for them in the last year.
Most reported reviewing \textit{more} than the guideline advocates (78), about a fourth roughly adhered to it (32), and very few were below it (10).
Unfortunately, 7 of the 10 respondents below the guideline were in tenured positions.
While some social-desirability bias is likely with such a question, the distribution suggests that the average ESE reviewer in our sample provides at least reciprocal reviews for their submissions.

\begin{figure}[h]
    \centering
    \includegraphics[width=\linewidth]{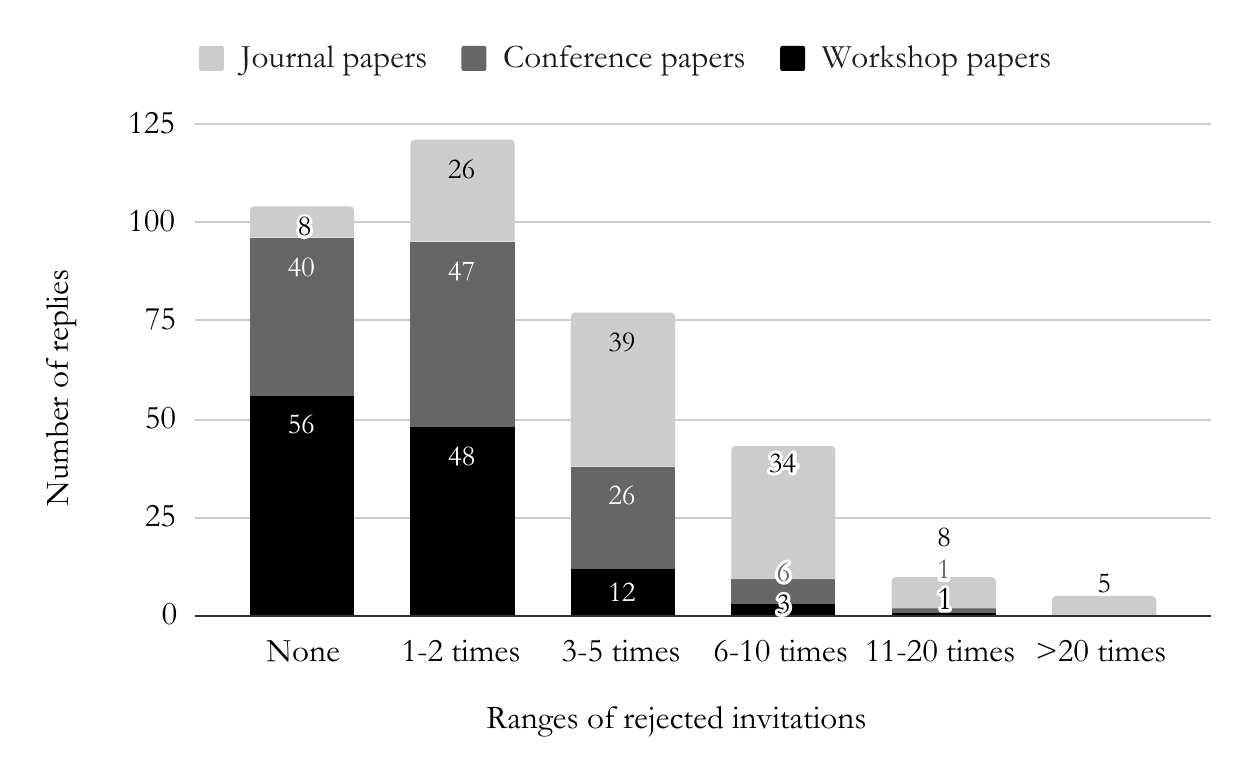}
    \caption{Distribution of declined review invitations across workshops, conferences, and journals in the last year}
    \label{fig:rejections}
\end{figure}

As another indicator for workload, we asked how often participants declined review invitations in the last year (see Fig.~\ref{fig:rejections}).
Workshop and conference invitations were seldom declined, with a median of \enquote{never} and \enquote{1-2~times} respectively.
Out of the 26 conference rejections in the \enquote{3-5~times} range, 21 were by tenured positions, who are naturally also more likely to receive PC invitations.
Review invitations by journal were generally declined more frequently, with 39 respondents in the \enquote{3-5~times} and 34 in the \enquote{6-10~times} range.
Given that journal reviews are individual requests compared to the PC invitations of conferences and workshops, the more frequent declines are expected here.
Nonetheless, it appears that journals are suffering more than other venue types from the current reviewer shortage.

As a final aspect of review load, the practice of sub-reviewing does not seem very widespread.
Most of the 120 respondents never delegated reviews (85) or invited sub-reviewers very rarely (22).
Only a minor portion relied moderately to often on sub-reviewers (10), and hardly any people reported always using sub-reviewers~(3).

\textbf{Our interpretation:} In summary, ESE reviewing workload is perceived as high and is mostly spent in conference PCs, a role that is more often covered by tenured academics.
Conference and workshop PC invitations are seldom rejected, and mostly by senior academic roles.
Declining journal review invitations is far more common.
A majority reports reviewing more than three papers for each submitted one.
Using sub-reviewers is not a widespread practice.
We could not find any specific data on how ESE research is spread out across workshops, conferences, and journals.
However, we speculate that the conference-centric nature of the data we collected is not due to the number of submitted ESE papers per venue type alone.
The observed trend might instead be influenced by the \enquote{invisible service} nature of journals.
Belonging to a conference PC provides visibility, supports building a traceable curriculum, and fosters a sense of community, all of which are harder to achieve with journal reviews.
This hypothesis seems to be supported by review invitation rejections: for journals, they are evenly spread out across all academic roles, while conference invitations are predominantly declined by established academics, who do not need the recognition of serving on another PC.

\section{Review Quality}
\begin{figure}[h]
    \centering
    \includegraphics[width=\linewidth]{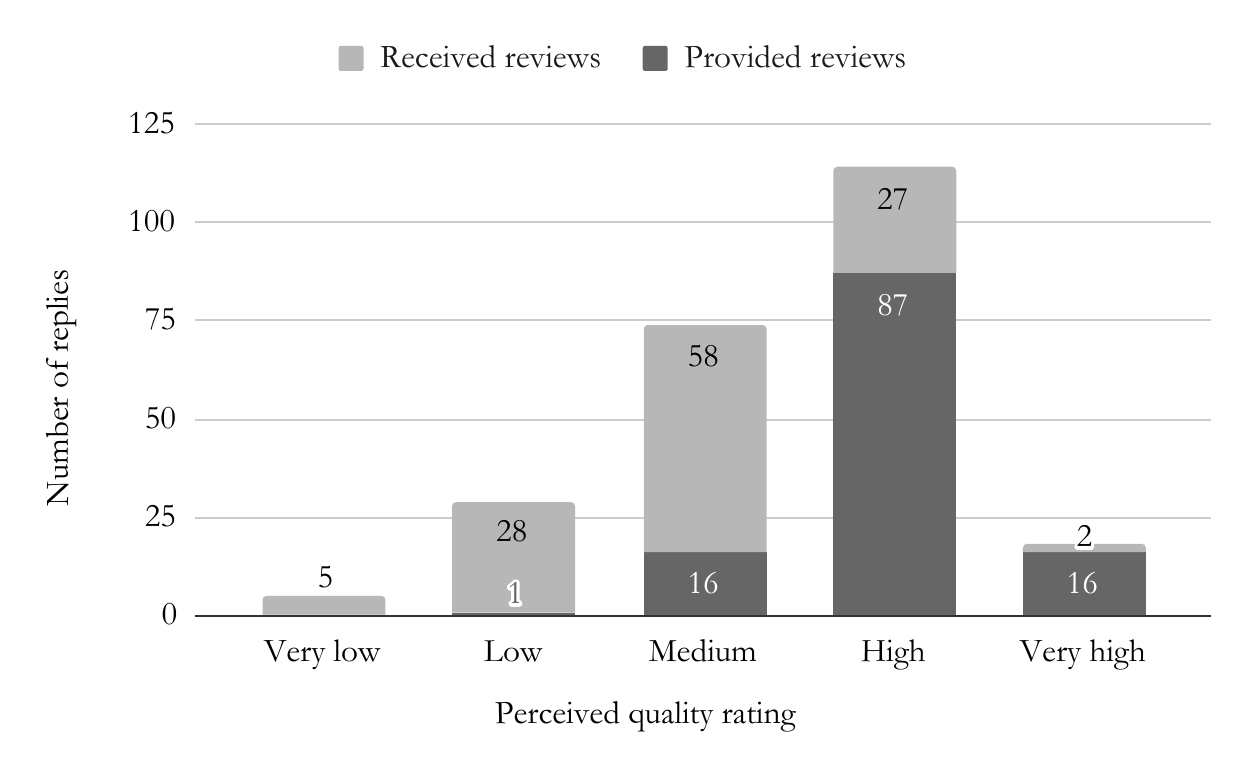}
    \caption{Perceived review quality in the last year}
    \label{fig:rev_quality}
\end{figure}

Regarding review quality, we asked ESE researchers to indicate the perceived quality of reviews provided by themselves and reviews received by others (see Fig.~\ref{fig:rev_quality}).
The vast majority of the 120 respondents considered their reviews to be of high (87) or very high (16) quality, while 16 selected medium and a single respondent low quality (no one chose \enquote{very low}).
Unsurprisingly, the distribution of received review quality shifted left: most participants reported at least medium (58), high (27), or very high (2) quality, but many more people now chose low (28) or very low (5).
The median difference between provided and received review quality was 1 point (4 vs. 3), which is expected to some degree.

\begin{figure}[h]
    \centering
    \includegraphics[width=\linewidth]{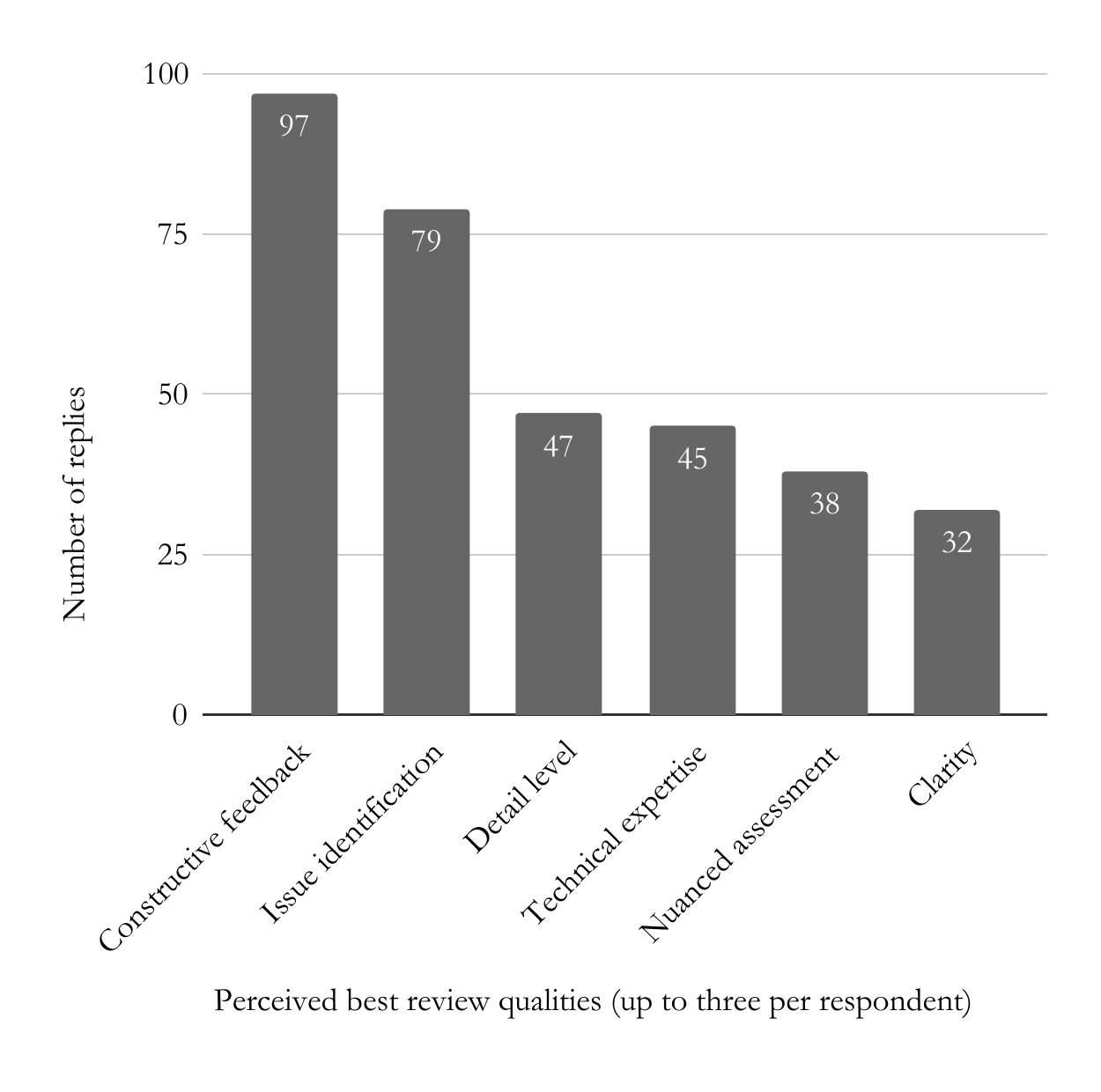}
    \caption{Perceived best qualities of provided reviews}
    \label{fig:rev_qualities}
\end{figure}

When asked what they perceive as the best qualities in their reviews, the participating ESE researchers provided different answers, but most shared two opinions (see Fig.~\ref{fig:rev_qualities}).
The vast majority of respondents (97) consider providing constructive and actionable feedback as one of their best review qualities, followed by the ability to identify key issues (79).
None of the other qualities were picked by the majority of participants, e.g., review thoroughness and level of detail (47) or knowledge of the topic \& technical expertise (45).
Interestingly, a nuanced \& balanced assessment was only selected by 37 participants (31\%), which may explain why there is so much anecdotal evidence of people perceiving reviews as unfair and harsh.
Lastly, more syntactic review qualities like clarity \& organization of the review also seem to be valued less (31).

\begin{figure}[h]
    \centering
    \includegraphics[width=\linewidth]{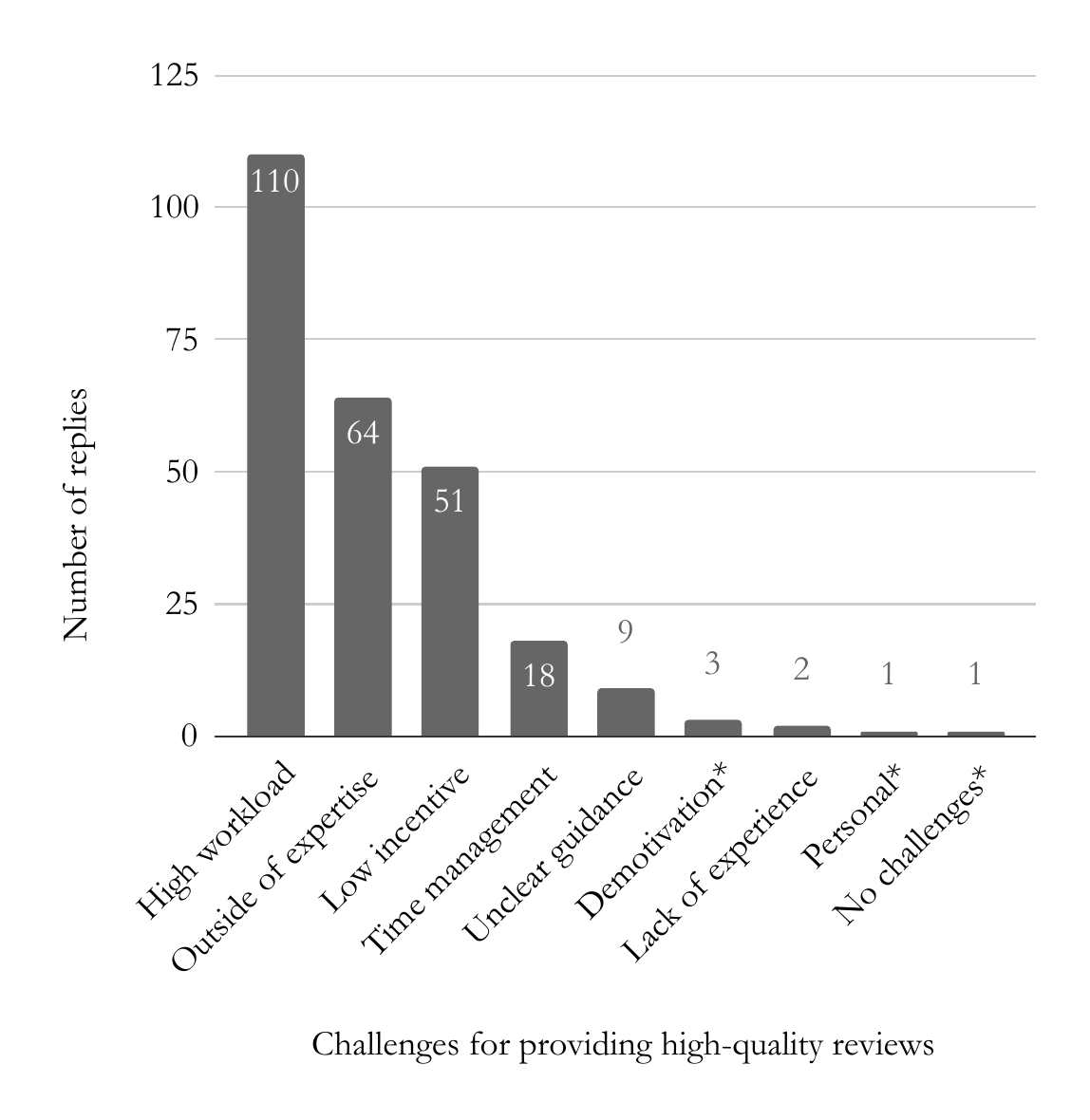}
    \caption{Most frequent challenges for providing high-quality reviews in the last year (coded \enquote{other:} options marked with an asterisk)}
    \label{fig:challenges}
\end{figure}

Regarding challenges that most frequently impede high-quality reviews (Fig.~\ref{fig:challenges}), the ESE community has a clear winner: we are overworked.
More than 90\% of respondents (110) perceived high workload and too many high-priority tasks as the main obstacle to providing higher-quality reviews.
Another frequently mentioned challenge was that assigned papers are too far from their expertise (64).
Given that more than half of participants reported this, there clearly are issues in how we currently assign reviews or in the expertise PCs and journals have available.
The third most frequent challenge was low personal incentives or insufficient recognition (51), which also means that we clearly can do better in this area.
Following these top 3 issues, fewer respondents reported procrastination and poor time management (18), unclear expectations and guidance from the venue (9), or a lack of experience and review tutoring (2).
From the custom \enquote{other} options, a few participants reported a sense of demotivation due to AI-generated, low-quality, or uninteresting papers (3).
Finally, one person reported personal circumstances (\enquote{young kids at home}) and another one mentioned that they do not experience any challenge.

\begin{figure*}[t]
    \centering
    \includegraphics[width=\linewidth]{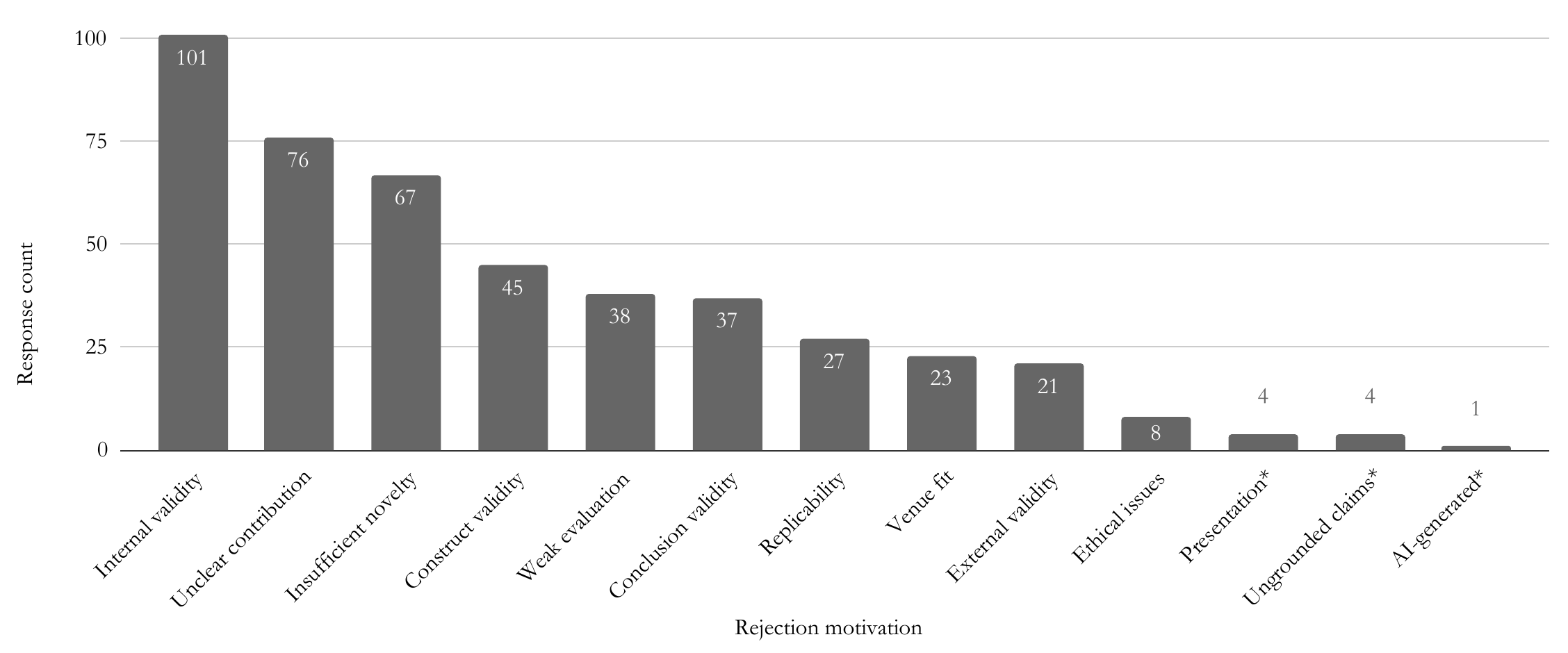}
    \caption{Most frequent reasons to argue for rejecting papers (coded \enquote{other:} options marked with an asterisk)}
    \label{fig:rationale}
\end{figure*}

When it comes to the most frequent reasons that lead reviewers to reject a paper (Fig.~\ref{fig:rationale}), most people reported insufficient methodological rigor, i.e., internal validity concerns (101).
While this aligns well with the ESE community's methodology focus, the second most frequent reason was slightly surprising: 76 people reported an unclear contribution or motivation, i.e., why is this research needed?
Combined with rank 3, namely insufficient novelty and/or positioning regarding related work (67), this may explain anecdotal evidence of people complaining about their methodologically sound studies being rejected.
Relevance and novelty are infamously difficult to assess and usually have a subjective nature.
Other issues related to research design, such as construct validity (45), insufficient evaluation of a design contribution (38), and insufficient or improper statistical analysis, i.e., conclusion validity (37), were also mentioned with decent frequencies.
About a fourth of respondents noted issues related to insufficient study replicability (27).
Less frequent mentions were relevance for and fit to the venue scope (23), insufficient generalizability, i.e., external validity (21), and ethical issues such as plagiarism and lack of consent (8).
As labeled \enquote{other} replies, participants reported presentation shortcomings (4) and ungrounded claims (4), with one person bringing up AI-generated content as the reason for rejection.
However, other participants may have included AI generation under ethical issues.

\begin{figure}[h]
    \centering
    \includegraphics[width=\linewidth]{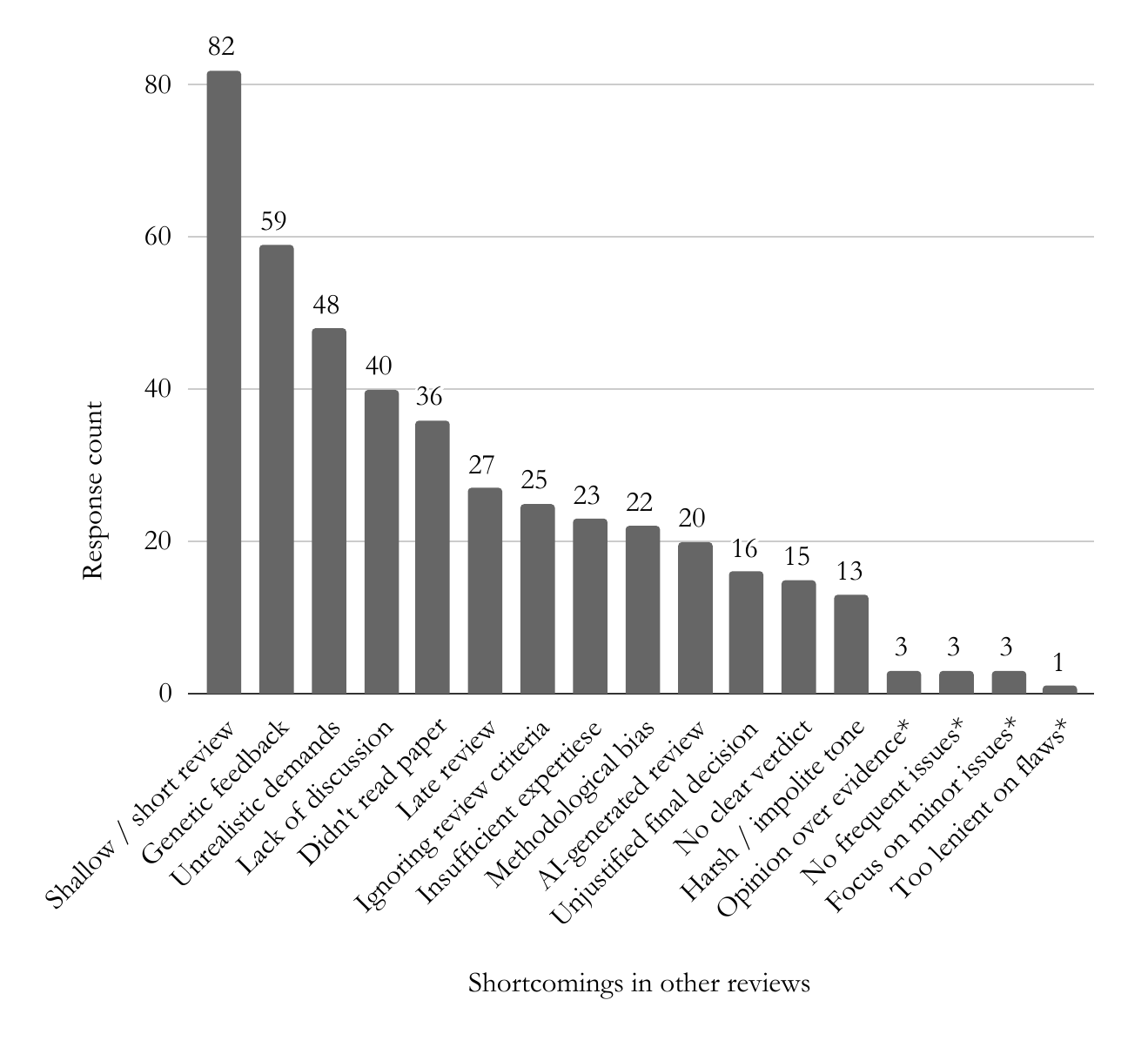}
    \caption{Most frequent issues noted in other reviews for the same paper (coded \enquote{other:} options marked with an asterisk)}
    \label{fig:other}
\end{figure}

As the final question regarding review quality and process, we asked what participants consider to be the most recurrent shortcomings of \textit{other} reviews for the same paper (see Fig.~\ref{fig:other}).
By far the most frequently reported issue was that reviews provided by others are shallow or very short (82), which seems consistent with the most frequent challenges of high workload and low incentives for high-quality, detailed reviews.
However, almost half of participants also reported generic, non-actionable feedback as a frequent issue in reviews (59), despite over 80\% claiming that providing constructive feedback is precisely one of their greatest strengths during reviewing.
Other somewhat frequent issues were unrealistic demands for additional work (48), a lack of discussion between reviewers (40), or late reviews by peers (27), which also makes discussion more difficult.
Less frequently mentioned issues were a lack of adherence to review criteria (25), a lack of reviewer expertise on the topic (23), and methodological biases by reviewers, e.g., strong opposition to qualitative methods (22).
While not among the top mentions, the issue of AI-generated reviews was already reported by a worrisome one-sixth of participants (20), which is also likely to increase in the near future.
Recurrent themes in the open-ended answers were about subjective review comments not grounded in evidence (3), an overemphasis of minor issues (3), or excessive leniency toward study design flaws (1).
The latter suggests that, on average, the ESE community does not seem to be in danger of being too lenient with paper acceptances.
Finally, three participants reported no relevant frequent shortcomings in other reviews, and we should definitely find out which venues they are typically a part of.

\textbf{Our interpretation:} Unsurprisingly, participants deemed the reviews they write of higher quality than those they receive, and several frequent issues are simultaneously among the reported top qualities of participants' reviews.
This points to either an above-average survey sample of diligent, high-quality reviewers, i.e., a self-selection bias of people truly interested in peer review, or a certain degree of self-assessment bias and/or social-desirability bias, e.g., when judging how actionable provided feedback really is.
A mix of the two is very likely, but it is difficult to assess with the data we have.
What is more consistent in the results is that reviewers experience a high workload, have low motivation to provide high-quality reviews due to missing incentives, and often review submissions outside their expertise.
As a consequence, reviews are often shallow or short, provide generic non-actionable feedback, make unrealistic demands, and lack proper  discussion between reviewers.
While AI-generated content is cited only once as a frequent rejection reason, the publish-or-perish pressure driving its unethical use in research may, as a defensive mechanism, be fueling a parallel rise in improper LLM use for reviews that is already ongoing (reported by 20 people), creating a vicious cycle with potentially damaging consequences for both the ESE community and science at large.
It will be important to break such a potential self-sustaining relationship early on, before it can spiral out of control.

\section{LLM Impact on Reviewing}
\begin{figure}[h]
    \centering
    \includegraphics[width=\linewidth]{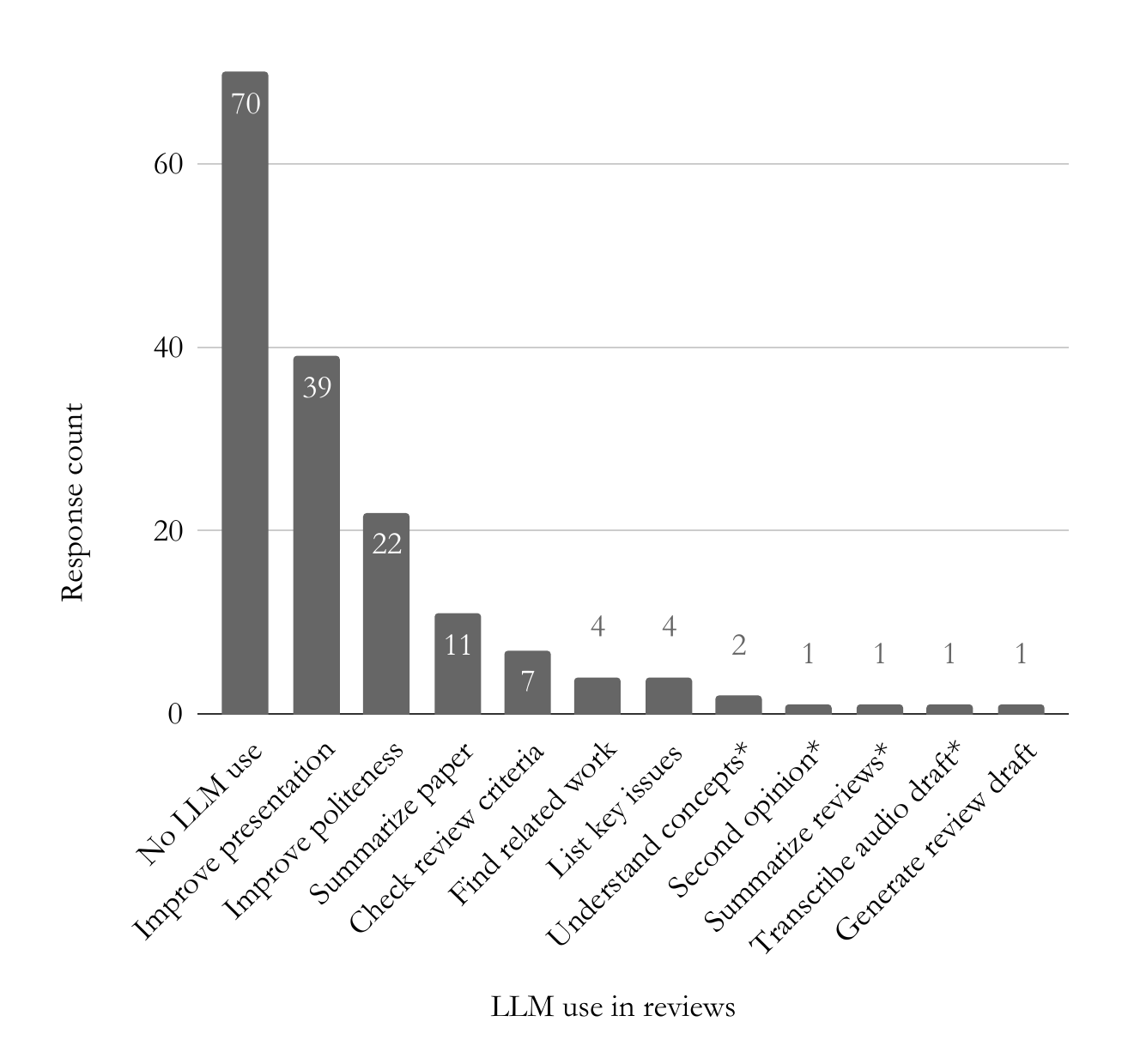}
    \caption{Self-reported LLM use during peer review (coded \enquote{other:} options marked with an asterisk)}
    \label{fig:llm-use}
\end{figure}

To understand how LLMs changed the ESE peer review process, we first asked participants in what capacity they use LLMs when reviewing (see Fig.~\ref{fig:llm-use}).
More than half of the 120 respondents do not use them at all (70).
The most frequently reported use cases were to improve review presentation quality (39) and to make reviews more respectful / polite (22).
Both of these do not require the LLM to process the paper, which is forbidden by most publisher and venue policies, at least for public GenAI services, as this would breach paper confidentiality.
Similarly reasonable use cases were checking if reviews adhere to the review instructions / evaluation criteria (7), identifying potentially missing related work for the paper topic (4), or understanding major concepts of the paper better (2).
However, several reported use cases also required the LLM to process the full paper and were therefore usually more ethically questionable, even if local LLMs were used.
For example, several people reported letting the LLM summarize the paper (11) or provide a list of key issues in the paper (4).
One person even reported letting the LLM generate a full review draft for manual refinement.

Regarding the used LLM or GenAI service, most users relied on ChatGPT (36), followed by Gemini (12) and Claude (7).
Other sporadic mentions were Perplexity (3), Microsoft Copilot (3), and DeepSeek (2).
Despite several reported use cases requiring complete processing of the paper, only six people used local LLMs to respect paper confidentiality.
Lastly, some people reported the usage of language-assistance tools like Grammarly (4) or DeepL (2) for this question.
Overall, the majority of LLM users seem to rely on public GenAI services, which, depending on the use case, can be dangerous regarding paper confidentiality.

\begin{figure*}[h]
    \centering
    \includegraphics[width=\linewidth]{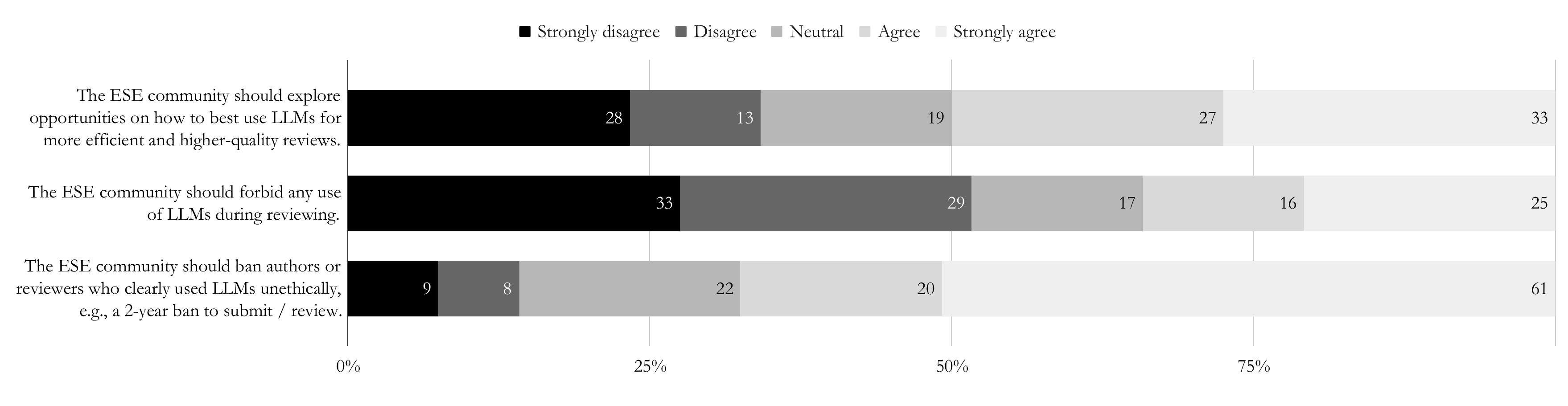}
    \caption{Opinions on LLM use in ESE peer review}
    \label{fig:likert-items}
\end{figure*}

Lastly, we wanted to hear participants' opinions about three statements regarding LLM use for peer reviews in the ESE community (see Fig.~\ref{fig:likert-items}), which were collected via 5-point Likert items (\enquote{strongly disagree} to \enquote{strongly agree}).
Regarding whether the ESE community should explore how to best use LLMs to support peer review, participants were divided.
While exactly half agreed or strongly agreed (60), 19 remained neutral, and 41 either disagreed or strongly disagreed (34\%).
We see a similar picture for whether the ESE community should completely forbid LLM use during reviewing.
Consistent with the previous question, 41 people agreed or strongly agreed (34\%), while 17 remained neutral, and 62 disagreed or strongly disagreed (52\%).
Overall, there seems to be a slight majority that wants to explore the responsible usage of LLMs to make peer review more effective and efficient, but we are very far from having broad consensus on this.
However, we do have consensus on one thing, namely whether we should ban authors and reviewers who use LLMs unethically.
A clear majority favors banning identified offenders (81 agree or strongly agree).
Only 22 people remained neutral, with 17 disagreeing or strongly disagreeing.

\textbf{Our interpretation:} So far, LLMs are used sparingly during ESE peer review and mostly for reasonable, ethically defensible activities.
However, it is likely that current usage is both more frequent and more questionable than reported by our selective sample of ESE researchers.
Overall, such questions are also impacted by social-desirability bias.
Still, in our sample, it seems probable that the vast majority of reviewers either do not use LLMs or only for small, reasonable tasks without major ethical consequences.
Nonetheless, it is definitely not ideal that the vast majority of LLM use relies on public GenAI tools without privacy guarantees.
If the ESE community wants to find responsible ways to integrate LLMs into peer review, we need to provide our own platforms for this that align with our values.
Currently, the community seems partially divided on whether we should start embracing LLMs during peer review, with a slight majority being in favor of exploring suitable use cases.
However, there is broad consensus to punish clear LLM misconduct of authors and reviewers.
While the policy details of what \enquote{unethical use} means may need some sharpening and discussion, the community seems to have had enough of letting offenders off the hook without repercussions.
This is in direct contrast to how, e.g., ICSE'26 handled GenAI-hallucinated references in accepted papers in the Research Track, allowing authors to correct them without any consequences instead of rejecting the papers and banning the authors.\footnote{\url{https://www.linkedin.com/posts/steffen-herbold-b2b4a854_the-icse-international-conference-on-software-share-7450164703143755777-1S6r}}

\section{Suggestions for Improvement}
\begin{figure}[hbpt]
    \centering
    \includegraphics[width=\linewidth]{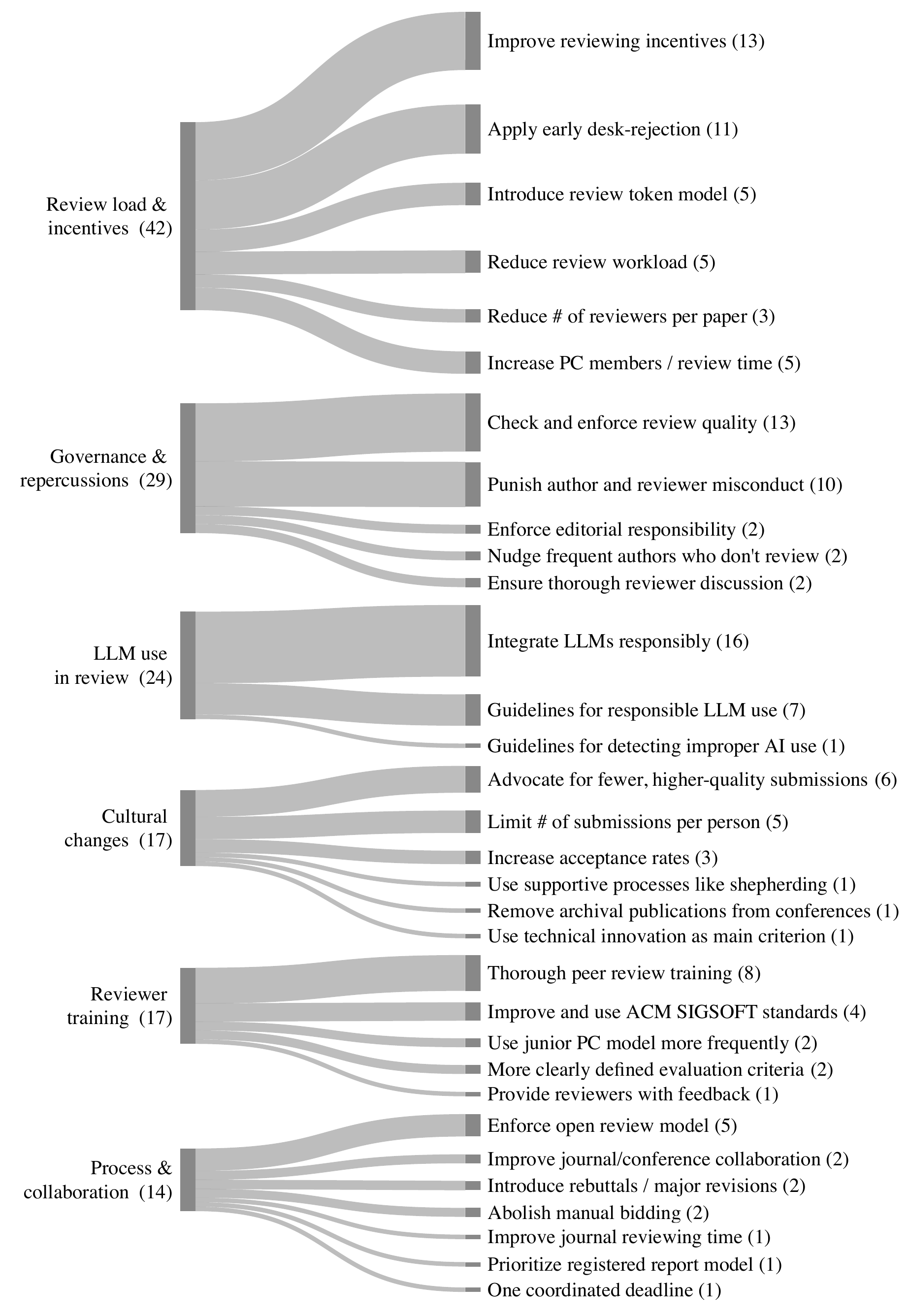}
    \caption{Coded open-ended suggestions on how to improve ESE review processes}
    \label{fig:suggestions}
\end{figure}

The last questionnaire section was about suggestions from the community to improve the current state of ESE peer review.
In total, 82 participants answered this optional free-text question, indicating the importance of the topic for the community.
We aggregated their feedback into 33 actions in 6 different categories (see Fig.~\ref{fig:suggestions}).

The largest category of suggestions centered around \textbf{review load and how to reduce it} (42).
For example, 13 people recommended improving reviewer incentives, e.g., by making the review contributions of individuals more transparent and visible (especially in relation to the number of submissions), by reducing or waiving conference fees for PC members, or by directly paying reviewers.
Another prominent suggestion (11) was to apply early desk rejection more extensively and to only forward promising papers to peer review.
Finally, people recommended introducing a review token model for submissions (5), e.g., as described by Amy J. Ko\footnote{\url{https://medium.com/bits-and-behavior/sustainable-peer-review-via-incentive-aligned-markets-a64ff726da56}}, reducing the number of reviewers per paper (3), e.g., from three to two, or increasing the number of PC members~(3).

The second largest group of suggestions was about \textbf{review governance and repercussions for misconduct} (29).
The most mentioned actions here were that chairs and editors should check and enforce review quality more thoroughly (13) and that serious punishments must be established for author and reviewer misconduct (10), e.g., 2-year bans on submitting and reviewing, which are also shared between venues.
Less frequently mentioned actions were to ensure thorough discussions between reviewers (2), even for journal reviewing, or to nudge authors who do not review enough for their number of submissions~(2).

Many suggestions also focused on the \textbf{use of LLMs during the review process} (24).
One frequently proposed action, in fact the single most mentioned one across all categories, was to responsibly integrate LLMs into the review process for improved quality and efficiency (16).
For example, some people proposed to provide AI-generated reviews by design next to human ones.
Others recommended using LLMs for the efficient pre-screening of papers or to check for GenAI-hallucinated references.
In addition to a systematic integration of LLMs into the review process, people also proposed to provide reviewers with guidelines on how to use LLMs responsibly (7) but also on how to detect improper AI use~(1).

Smaller categories were about \textbf{cultural changes} (17), e.g., advocating for a culture of fewer, higher-quality submissions (6) in combination with limiting the number of submissions per person (5) and increasing acceptance rates (3); \textbf{reviewer training and guidance} (17), e.g., establishing thorough peer review training (8), improving and using the ACM SIGSOFT empirical standards (4), or using the junior PC model more frequently (2); and \textbf{process changes and collaboration} (14), e.g., enforcing the open review model (5) similar to what AIware'25 introduced\footnote{\url{https://2025.aiwareconf.org/\#openreview}}, improving collaboration and sharing between journals / conferences (2) similar to what the EiCs of the major SE journals proposed~\cite{Menzies2026}, and to introduce rebuttals and/or major revisions into all conferences~(2).

\textbf{Our interpretation:} In summary, proposals from the ESE community to reduce review load and increase review quality cover bringing in more people via effective incentives, making sure everyone provides their fair share, punishing misconduct and abuse of the system, checking and enforcing review quality, and finding responsible ways to integrate LLMs to lighten the review load.
Interestingly, apart from the responsible LLM use for peer review, the vast majority of suggestions seem rather conservative, more like thoroughly and extensively applying mechanisms we or other scientific communities already have (partially) in place, e.g., banning offenders, larger PCs, fewer reviews per paper, checking review quality, junior / shadow PCs, more time to review, or more early desk rejects.
More extensive or radical proposals that require a cultural change or a fundamental rework of the system were fairly rare, e.g., advocating for a culture of fewer, higher-quality submissions instead of our current publication madness.
While five participants advocated for the slightly more profound change of a review token model, only one person dared to suggest abolishing archival publications from our conferences, which would finally align us with many other fields.
Lastly, not a single participant commented in the direction of reducing or abolishing (pre-publication) peer review, something that other academic communities have already considered and discussed for years~\cite{Heesen2021}.

\section{Conclusion}
Our survey results clearly show that peer reviewers in the ESE community are under high review load and that something needs to be done: the current state of affairs is clearly not sustainable anymore and likely has not been for a while.
The reported average received review quality could, of course, be much worse, but we still can do better than this, especially considering the many reported frequent issues in other reviews.
While some participants seem carefully optimistic about making good use of LLMs to improve peer review, many other responses, especially the many free-text comments, also tell a story of disappointment, anger, frustration, and uncertainty.
Comments like \enquote{I'm honestly somewhat dis-illusioned about the state of peer review in science. I am not sure how to address this.}, \enquote{I am not going to accept doing reviews in 2026 unless I'm fascinated by the topic, and others will no doubt start doing the same. The situation is untenable.}, or \enquote{One of my PhD student is starting to be negatively affected by the fact that in almost all our submissions we had at least one very bad review.} clearly show that many people have had enough.
Change is needed, and it is needed fast.
While our community may still do better than others, we can see a preview of what might happen if we do not act soon when looking at the AI communities.
For example, hallucinated references were identified in over 50 accepted papers at NeurIPS'25~\cite{Ansari2026} and 21\% of reviews of ICLR'26 were very likely AI-generated~\cite{Naddaf2025}.

While many of the proposed solutions seem reasonable, one issue around peer reviewing is that, for a mechanism that is so central to our work, we still have comparatively little empirical evidence of what effective and efficient peer review is supposed to look like~\cite{Aczel2025}.
Additional research would certainly be helpful to guide us.
Whatever we decide as a community, it will be important to accompany any introduced changes with trustworthy evaluations about their effects, both positive and negative, so that we can adjust course if necessary.
However, in addition to these solutions to fight the symptoms, we should not forget major root causes, like our \enquote{publish or perish} culture, which also needs fixing.
Especially senior members of the community will need to start effecting change in this direction because one thing is clear: we cannot expect the next generation of PhD students to fix this for us, and our changes also cannot be to their detriment.

\section*{Acknowledgements}
We kindly thank Daniel Graziotin (University of Hohenheim), Patricia Lago (VU Amsterdam), Ivano Malavolta (VU Amsterdam), and Marvin Wyrich (Saarland University) for providing feedback during the pilot study. Additionally, we thank our 120 survey participants for their valuable time and insights.

\bibliographystyle{plainnat}
\bibliography{biblio}

\end{document}